\documentstyle[12pt]{article}
\newcommand{\be}{\begin{equation}}
\newcommand{\bea}{\begin{eqnarray}}
\newcommand{\eea}{\end{eqnarray}}
\newcommand{\ba}{\begin{array}}
\newcommand{\ea}{\end{array}}
\newcommand{\ee}{\end{equation}}

\expandafter\ifx\csname mathbbm\endcsname\relax

\else

\fi
\def\l{\label}
\def\o{\over}

\begin{document}
\begin{titlepage}
\hfill
\vbox{
    \halign{#\hfil         \cr
           hep-th/9609157 \cr
           IPM-96-165   \cr
           Sep 1996   \cr
           } 
      }  
\vspace*{3mm}
\begin{center}
{\LARGE On the Picard-Fuchs equations of the SW models \\}
\vspace*{20mm}
{\ Mohsen Alishahiha \footnote{e-mail:alishah@physics.ipm.ac.ir}}\\
\vspace*{1mm}
{\it Institute for Studies in Theoretical Physics and Mathematics, \\
 P.O.Box 19395-5531, Tehran, Iran } \\
{\it Department of Physics, Sharif University of Technology, \\
\it  P.O.Box 11365-9161, Tehran, Iran }\\
\vspace*{25mm}
\end{center}
\begin{abstract}
We obtain the closed form of the Picard-Fuchs equations for $N=2$ 
supersymmetric Yang-Mills theories with classical Lie gauge groups. 
For a gauge group of rank $r$, there are $r-1$ regular and an exceptional 
differential equations. We describe the series solutions 
of the Picard-Fuchs equations in the semi-classical regime.
\end{abstract}
\end{titlepage}
\newpage
From duality and holomorphy, Seiberg and Witten\cite{SW} have obtained the
exact prepotential of $N=2$ SYM theory with gauge group $SU(2)$ by studying the
singularities  of its moduli space at strong coupling. 
For a gauge group of rank $r$, the Seiberg and Witten's 
data is a hyperelliptic curve with $r$ complex dimensional moduli space
with certain singulareties and a meromorphic one form, ($E_{u_i},\lambda_{SW}$).
More precisely, the prepotential of $N=2$ SYM in the Coulomb phase can be 
described
with the aid of a family of complex curves with the identification of the
v.e.v.'s $a_i$ and their duals $a^{D}_{i}$ with the periods of the curve
\be
a_{i} = \oint _{\alpha _{i}} \lambda_{SW} \;\;\;\; \mbox{and} \;\;\;\;
a^{D}_{i} = \oint _{\beta_{i}} \lambda_{SW} , \label{integrals}
\ee
where $\alpha_{i}$ and $\beta_{i}$ are the homology cycles of the
corresponding Riemann surface.

To find the periods one should calculate the above integrals, or
one may use the fact that the periods $\Pi=(a_i,a^{D}_i)$
satisfy the Picard-Fuchs equations. So it is
important to find the Picard-Fuchs equations which also help
in the instanton calculus.

In this letter we obtain closed forms for the Picard-Fuchs equations for 
classical Lie gauge groups. Recently, some of these equations have been 
obtained in {\cite{IS} and \cite{KL}}.

The Seiberg-Witten's data ($E_{u_i},\lambda_{SW}$) for classical gauge groups
are known \cite{AA} and \cite{DH}, 
and take the following form
\bea  \l{CU}
y^2&=&W^2(x)-\Lambda ^{2\hat h}x^{2k}\cr
\lambda_{SW}& =&(k W-x{dW \o dx}){{dx} \o y}
\eea
where $\hat h$ is the dual Coxeter number of the Lie gauge group and 
\be \l{PW}
W(x)=x^m -\sum_{i=2}^{m}{u_i x^{m-i}}
\ee
with $m=r+1, i=2,3,...,r+1$ for $A_r$ series and $m=2r, i=2,4,...,2r$ for
$B_r, C_r,D_r$ series, and $u_i$'s, the Casimirs of the gauge groups. 
Also $k=m-\hat h$. Note that the $D_r$ series has
an exceptional Casimir, $t$, of degree $r$, but in our notation we set 
$u_{2r}=t^2$.

From explicit form of $\lambda_{SW}$ and the fact that the $\lambda_{SW}$
is lineary dependent on Casimirs, setting ${\partial \o {\partial{u_i}}}=
\partial_i$ we have
\bea \l{PARTIAL}
\partial_i\lambda_{SW}&=&-{x^{m-i} \o y}dx+d(*), \cr
\partial_i\partial_j\lambda_{SW}&=&-{x^{2m-i-j} \o y^3}W(x)dx+d(*).
\eea

By direct calculation one can see that
\be \l{DI}
{d \o dx}({x^n \o y})=(n-k){x^{n-1} \o y}+({{kx^{n-1} W-x^n{dW \o dx}} \o y^3})
W.
\ee
By inserting equation (\ref{PW}) in (\ref{DI}) we have
\be \l{DIF}
{d \o dx}({x^n \o y})=(n-k){x^{n-1} \o y}-{\hat h}{x^{m+n-1} \o y^3}W+ 
\sum_{i=2}^{m}{(m-k-i)u_i{x^{m+n-1-i} \o y^3}W}
\ee
Now from equations (\ref{PARTIAL}) we can find the second 
order differential equation for the periods $\Pi$ as follows  
\be  \l{PF}
{\cal L}_{n}=(k-n)\partial_{m-n+1}+{\hat h}\partial_2\partial_{m-n-1}-
\sum_{i=2}^{m}{(m-k-i)u_i \partial_i\partial_{m-n+1}}.
\ee
where $n=s-1$ for $A_r$ series and $n=2s-1$ for $B_r,C_r$
and $D_r$ series and $s=1,...,r-1$. Note that in the $A_r$ series, for $s=1$, the
above expression is not valid. In fact we should be careful in the final step 
in the derivation of the equation (\ref{PF}), that ${\cal L}^{A_r}_0$ is
\be  \l{APF}
{\cal L}^{A_r}_0=(r+1)\partial_2\partial_r-\sum_{i=2}^{r}{(r+1-i)u_i 
\partial_{i+1}\partial_{r+1}}
\ee
Moreover for $s>r-1$ equation (\ref{DIF}) does not give
the second order differential equation with respect to $u_i$. So, by this 
 method we can only 
find $r-1$ equations which we call {\it regular} equations. 
Also from the equation (\ref{PARTIAL}) we have the following identity 
\be\ba {ll} \l{ID}
{\cal L}_{i,j;p,q}=\partial_i\partial_j-\partial_p\partial_q, & i+j=p+q
\ea\ee

The $r$th equation, the {\it exceptional} equation, can be obtain from 
the following linear combination
\be
D=(k-m)d({x^{m+1} \o y})+\sum_{i=2}^{m}{(m-k+i)u_i
d({x^{m+1-i} \o y})}
\ee
or
\be
D=\lambda_{SW}-(\sum_{i=2}^{m}{i(i-2)u_i {x^{m-i} \o y}}
+\sum_{j,i=2}^{m}{ij u_iu_j {x^{2m-i-j} \o y^3}W} -{\hat h}^2 
\Lambda^{2{\hat h}} {x^{2k} \o y^3}W)dx.
\ee

So from the equation (\ref{PARTIAL}), the {\it exceptional} differential
equation for the periods $\Pi$ are 
\be
{\cal L}_r=1+\sum_{i=2}^{m}{i(i-2)u_i\partial_i}+\sum_{j,i=2}^{m}{ij u_iu_j
\partial_i\partial_j}-{\hat h}^2 \Lambda^{2{\hat h}}\partial_{{\hat h}}^{2}.
\ee
for $A_r,D_r$ and $C_r$ with odd $r$. For $B_r$ and $C_r$ with even $r$
the last term should be changed 
to $-{\hat h}^2 \Lambda^{2{\hat h}}\partial_{{\hat h}-1}\partial_{{\hat h}+1}$.

This equation together with the equations (\ref{PF}) and (\ref{ID})
give a complete set of the Picard-Fuchs equations for the periods 
$(a_i,a^{D}_i)$.

To study the series solutions of the Picard-Fuchs equations in the 
semi-classical regime, let us rewrite the Picard-Fuchs equations in 
terms of Euler derivative $\vartheta_i=u_i\partial_i$. 
The regular equations for $s\neq r-1$
become
\bea
{\cal L}_n=[(k+1-n)\vartheta_{m-n+1}-1]{\vartheta_{m-n+1} \o u_{m-n+1}}-
\sum_{{i\neq m-n+1},i=2}^{m}{{(m-k-i)\o u_{m-n+1}}\vartheta_i\vartheta_{m-n+1}}\cr
+{{\hat h}\o u_2 u_{m-n-1}}\vartheta_2\vartheta_{m-n-1} 
\eea
If $s=r-1$ the last term should be replaced by ${{\hat h}\o u_2^2}
\vartheta_2(\vartheta_2-1)$. Again, we should be careful for $s=1$ in the
$A_r$ series. From equation (\ref{APF}) we have 
\be
{\cal L}_0^{A_r}=
{(r+1)\o u_2 u_r} \vartheta_2\vartheta_r-{u_r \o u_{r+1}^2}
\vartheta_{r+1}(\vartheta_{r+1}-1) -\sum_{i=2}^{r-1}{{(r+1-i)u_i \o u_{i+1}
u_{r+1}}\vartheta_{i+1}\vartheta_{r+1}}
\ee
The exceptoinal equation changes to
\be
{\cal L}_r=
(1-\sum_{i=2}^m{i\vartheta_i})^2-{{\hat h}^2\Lambda^{2{\hat h}}\o u_{{\hat h}^2}} 
\vartheta_{{\hat h}}(\vartheta_{{\hat h}}-1)
\ee
for $A_r,D_r$ and $C_r$ with odd $r$, and 
\be
{\cal L}_r=
(1-\sum_{i=2}^m{i\vartheta_i})^2-{{\hat h}^2\Lambda^{2{\hat h}}\o u_{{\hat h}-1}
u_{{\hat h}+1}} 
\vartheta_{{\hat h}-1}\vartheta_{{\hat h}+1}
\ee
for $B_r$ and $C_r$ with even $r$. 

Let us define the variables $x_s$  
\be\ba {ll}
x_s={u_{m-n+1} \o {u_2 u_{m-n-1}}} & s=1,...,r-1 \\
& \\
x_r={\Lambda^{2{\hat h}} \o u_{{\hat h}}^2}& (or \,\,\,
x_r={\Lambda^{2{\hat h}} \o {u_{{\hat h}-1}u_{{\hat h}+1}}})
\ea\ee
for $B_r, C_r$ and $D_r$ corresponding to above, and the variables
\be\ba {ll}
x_s={u_{m-n+1} \o {u_2 u_{m-n-1}}} & s=2,...,r-1 \\
&\\
x_1={\Lambda^{2(r+1)} \o u_2u_{r}^2}&x_r={\Lambda^{2(r+1)} \o u_{(r+1)}^2}.
\ea\ee
for the $A_r$ series.  
We can construct the power series solution of the Picard-Fuchs equations
around $(x_i)=(0)$ \cite{IS}
\be \l{SUM}
\omega(a_1,...,a_r;x_1,...,x_r)=\sum_{l_1,...,l_r=0}{C_{l_1,...,l_r}
x_1^{l_1+a_1}...x_r^{l_r+a_r}}
\ee

Let us take $\alpha_i(l_j)=\alpha_i(a_1,...,a_r;l_1,...,l_r)$ be 
the power of $u_i$ when equation (\ref{SUM}) is reexpressed in term  
of $u_i$'s. By inserting $\omega$ in the Picard-Fuchs equations, 
one can obtain the indicial and recursion relations. 
For example the indicial relation for $B_r,C_r$ and $D_r$ are
\bea
[(k+1-n)\alpha_{m-n+1}(0)-1]\alpha_{m-n+1}(0)&-&
\sum_{{i\neq m-n+1},i=2}^{m}{(m-k-i)\alpha_i(0)\alpha_{m-n+1}(0)}=0 \cr
(1-\sum_{i=2}^{m}{i\alpha_i(0))^2}&=&0 
\eea

Also the recursion relations are
\be\ba {ll} \l{RE}
C_{l_1,...,l_r}={{-{\hat h}\alpha_2(l_s-1)\alpha_{m-n-1}(l_s-1)}\o
\Delta_s}C_{l_1,...,l_s-1,...,l_r} & i=1,...,r-2\\
& \\
C_{l_1,...,l_r}={{-{\hat h}\alpha_2(l_{r-1}-1)(\alpha_2(l_{r-1}-1)-1)}\o
\Delta_s} C_{l_1,...,l_{r-1}-1,l_r} & \\
  &   \\
C_{l_1,...,l_r}={{{\hat h}^2\alpha_{{\hat h}}(l_r-1)(\alpha_{{\hat h}}(l_r-1)-1)
} \o (1-\sum_i{i\alpha_i(l_j))^2}}C_{l_1,...,l_r-1}&
\ea\ee
where $\alpha_i(0)$ means that all $l_j=0$ and $\alpha_i(l_s-1)$ means that
the $s$th $l$ should be set equal to $l_s-1$ and other $l$'s are fixed, also
\be
\Delta_s= 
[(k+1-n)\alpha_{m-n+1}(l_j)-1]\alpha_{m-n+1}(l_j)-
\sum_{{i\neq m-n+1},i=2}^{m}{(m-k-i)\alpha_i(l_j)\alpha_{m-n+1}(l_j)}
\ee

Note that in the last relation of the equation (\ref{RE}) the suitable 
change as noted above should be made. By the same 
method one can obtain the indicial and recursion relations for the 
$A_r$ series. This method can be applied for the $N=2$ theories with
massless hypermultiplets which their curves are in the form of (\ref{CU}).

After completion of this work, i recived the paper \cite{MU} which is paid
to the same problem.
\vspace*{5mm}

I would like to thank M. Khorrami for helpful discussions.


\newpage
\begin{thebibliography}{99}
\bibitem{SW}
N. Seiberg and E. Witten, Nucl. Phys. B426 (1994) 19.

N. Seiberg and E. Witten, Nucl. Phys. B431 (1994) 484.

\bibitem{IS}
K. Ito, N. Sasakura hep-th/9608054

\bibitem{KL}

A. Klemm, W. Lerche, and S. Theisen,
Int. J. Mod. Phys. A11 (1996) 1929.

K. Ito, S. K. Yang hep-th/9603073.

Y. Ohta, hep-th/9604051, hep-th/9604059.

M. Matone, Phys. lett. B357 (1995) 342.

H. Ewen, K. F\"orger, S. Theisen, hep-th/9609062.

\bibitem{AA}

M. R. Abolhasani, M. Alishahiha, A. M. Ghezelbash, hep-th/9606043.

A. Klemm, W. Lerche, S. Yankielowicz and S. Theisen,
Phys. Lett. B344 (1995) 169.

P. Argyres, A. Faraggi, Phys. Rev. Lett. 73 (1995) 3931.

U. H. Danielsson, B. Sundborg, Phys. Lett. B358 (1995) 273.

A. Brandhuber, K. Landsteiner, Phys. Lett. B358 (1995) 73.

P. C. Argyres, A. D. Shapere, Nucl. Phys. B461 (1996) 437.

\bibitem{DH}

E. D'Hoker, I. M. Keichever, D. H. Phong, hep-th/9609145

\bibitem{MU}

J. M. Isidro, A. Mukherjee, J. P. Nunes, H. J. Schnitzer, hep-th/9609116.
\end{thebibliography}
\end{document}